\documentclass[aps,prd,preprint,groupedaddress,showpacs,floatfix]{revtex4-1}

\usepackage{epsfig}
\usepackage{graphicx}
\usepackage{amsmath}
\usepackage{amsfonts,amssymb}
\usepackage{amsfonts}
\usepackage{color}

  \usepackage{slashed}
  \usepackage{url}
\usepackage{bm}
\usepackage{verbatim}
\usepackage{float}

\def\endprf{\hfill  {\vrule height6pt width6pt depth0pt}\medskip}

\begin{document}


\title{Einstein-Maxwell equations for asymmetric resonant cavities}


\author{Marco Frasca}
\email[]{marcofrasca@mclink.it}
\affiliation{Via Erasmo Gattamelata, 3 \\ 00176 Roma (Italy)}


\date{\today}

\begin{abstract}
We analyze the behavior of electromagnetic fields inside a resonant cavity by solving Einstein--Maxwell field equations. It is shown that the modified geometry of space-time inside the cavity due to a propagating mode can affect the propagation of a laser beam. It is seen that components of laser light with a shifted frequency appear originating from the coupling between the laser field and the mode cavity due to gravity. The analysis is extended to the case of an asymmetric resonant cavity taken to be a truncated cone. It is shown that a proper choice of the geometrical parameters of the cavity and dielectric can make the gravitational effects significant for an interferometric setup. This could make possible to realize table-top experiments involving gravitational effects. 
\end{abstract}


\maketitle


\section{Introduction}

A single plane wave always induces a deformation of the geometry of the space-time \cite{Misner:1974qy}. This effect is so small and plane waves such idealized objects that hopes to observe it are certainly very tiny. Anyway, electromagnetic fields are easily available and the technology is old and it is not impossible to realize devices where the intensity of such fields could make this gravitational effect observable. This entails a rather sensible interferometer but it is not impossible to realize. The devices that better fit the aim are resonant cavities where, due to large merit factors, gross intensity of the electromagnetic energy can be achieved. The observation of such an effect would mean a real breakthrough in experimental general relativity as, so far, only large scale measurements were considered possible and available so far. A table-top experiment would completely change our way to perform test of general relativity in laboratory.

In order to approach this kind of physical problems we have to manage Einstein-Maxwell equations \cite{Misner:1974qy}. This is generally a rather involved task but the smallness of the effects we are going to study makes possible the application of perturbation techniques.

In this paper I show as a resonant cavity, with a single mode excited inside, could provide a satisfactory set-up for a measurement of the deviation of the geometry of the space-time from the flat case. In this case we have a laser beam traversing the resonant cavity. The effect of the gravitational field is to couple it with the modes inside the resonant cavity providing satellite frequencies, with respect to the laser frequency, that should be observed in the output. This could explain recent measurements done at NASA Eagleworks. It should be considered just the starting point for a more extended treatment to such an experimental set-up, much on the same lines of Ref.\cite{Minotti:2013mxa}. In the latter paper a modified Einstein-Maxwell theory involving scalar fields was considered but, with proper adjustments, but our aim is just to evaluate the effects at work through Einstein-Maxwell equations without further modification. In this way one can gets an estimation of the orders of magnitude involved.

Then, I analyze the case of a cavity having the form of a truncated cone. I show that, with a proper choice of the geometrical parameters of the cavity and dielectric, gravitational effects could be enhanced. We are able to evaluate the reaction of the gravitational field, a thrust computed in this particular geometry, and it is shown that such a cavity is best suited for enhancing such an effect due to the interplay of electromagnetic field, space-time geometry and geometry of the resonant cavity. 

The paper is so structured. In Sec.~\ref{secI} we introduce the Einstein-Maxwell equations. In Sec.~\ref{secII} we analyze the case of a plane wave applied to a resonant cavity in a form of a box. We assume then that an interferometric experiment is performed making a laser beam pass through the cavity. In Sec~\ref{secIII} we consider a more general cavity having the form of a truncated cone. We assume the approximation that the major radius is taken much greater than the minor radius while the latter is taken going to zero. This geometry proves to be favorable to the emergence of gravitational effects at observational level, even if really small. In Sec.~\ref{secIV} conclusions are presented.

\section{Einstein-Maxwell equations\label{secI}}

The set of Einstein-Maxwell equations is yielded by
\begin{equation}
    R^{\mu\nu}-\frac{1}{2}g^{\mu\nu}R=\kappa T^{\mu\nu}
\end{equation}
being $\kappa=8\pi G/c^4$ with $G$ the Newton constant and $c$ the speed of light, $g^{\mu\nu}$ the metric tensor and $R^{\mu\nu}$ the Ricci tensor with $R=R^\mu_\mu$. 
The energy-momentum tensor is given by
\begin{equation}
    T^{\alpha \beta} = \frac{1}{\mu_0} \left( F^{\alpha}{}^{\psi} F_{\psi}{}^{\beta} - \frac{1}{4} g^{\alpha \beta} F_{\psi\tau} F^{\psi\tau}\right)
\end{equation}
being the electromagnetic field tensor
\begin{equation}
    F^{\alpha\beta}=A^{\alpha;\beta}-A^{\beta;\alpha}=A^{\alpha,\beta}-A^{\beta,\alpha}
\end{equation}
where $A^\alpha$ is the vector potential, $;$ implies covariant derivative and $,$ is ordinary partial derivative. $F^{\alpha\beta}$ must satisfy the Maxwell equations
\begin{equation}
    F^{\alpha\beta}_{;\rho}=0
\end{equation}
without sources. It is also
\begin{equation}
    F^{\alpha\beta}_{;\delta}+F^{\delta\alpha}_{;\beta}+F^{\beta\delta}_{;\alpha}=0
\end{equation}
that completes the full set of Maxwell equations. These equations can be recast into the exact form \cite{Thorne:1980ru}
\begin{equation}
    -\frac{1}{2}\Box\overline{h}^{\mu \nu }=\kappa (T^{\mu \nu }+\tau^{\mu\nu}_{LL})
\end{equation}
being
\begin{equation}
    h^{\mu\nu}=g^{\mu\nu}-\eta^{\mu\nu}
\end{equation}
and $\tau^{\mu\nu}_{LL}$ the Landu-Lifshitz tensor \cite{LL2,Thorne:1980ru,Wein}. We note that $h^{\mu\nu}$ and $\eta^{\mu\nu}$ are not real tensors and we agree to raise or lower indexes of $h^{\mu\nu}$ by $\eta^{\mu\nu}$, the flat metric. Given the tensor $\overline h^{\mu\nu}=h^{\mu\nu}-\eta^{\mu\nu}h$ with $h=\eta_{\mu\nu}h^{\mu\nu}$, we choose the Donder gauge $\overline h^\mu_{\nu,\mu}=0$. This is the form we use in this article to perform perturbation theory. This is the standard setup for weak field approximation that is also the case we are interested in.

\section{Plane wave geometry\label{secII}}

\subsection{Geometry}

The simplest case discussed in literature for the Einstein-Maxwell equations is that of a plane wave \cite{Misner:1974qy}. I take the metric in the form
\begin{equation}
    ds^2=L^2(v)(dx^2+dy^2)-dvdu
\end{equation}
given the Rosen coordinates $v=ct-z$ and $u=ct+z$. It is easy to show that an electromagnetic plane wave modifies the geometry of space-time. I have that the Einstein tensor reduces to the Ricci tensor as the trace of the energy-momentum tensor is zero in this case. I will have the only non-null component
\begin{equation}
    R_{33}=-2\frac{L''(v)}{L(v)}.
\end{equation}
The electromagnetic field tensor will have the non-null components
\begin{equation}
    F_{01}=-F_{31}=A'(v).
\end{equation}
So, the only nonzero component of the energy-momentum tensor is
\begin{equation}
    T_{33}=-\frac{1}{\mu_0}\frac{(A'(v))^2}{L^2(v)}
\end{equation}
and so I have to solve the equation
\begin{equation}
    2\frac{L''(v)}{L(v)}=-\frac{8\pi G}{c^4\mu_0}\frac{(A'(v))^2}{L^2(v)}
\end{equation}
That has the solution $L(v)=\pm \alpha A'(v)$ provided
\begin{equation}
    \alpha^2=\frac{4\pi G}{c^2\omega^2\mu_0}
\end{equation}
and I am left with the equation for a plane wave
\begin{equation}
    [A'(v)]''+\frac{\omega^2}{c^2} A'(v)=0
\end{equation}
for the electromagnetic field and taking $A'(0)=E_0/c$ the magnetic field amplitude. Note that $\alpha\approx 9\cdot 10^{-21}\ A\cdot m\cdot N^{-1}=9\cdot 10^{-21}\ T^{-1}$ for $\omega=1\ GHz$. This is a small number as expected and this effect is negligible small for all practical purposes. Its inverse identify a critical magnetic field for which this effect could be meaningful but has an unphysical large value. Though it can be seen that, by a proper choice of the parameters entering into the definition of $\alpha$, this effect could nbe made visible with an interferometric technique.

In a resonant cavity, an estimation of the amplitude of the electric field $E_0$ can be computed using the formula \cite{Minotti:2013mxa}
\begin{equation}
    \frac{\epsilon_0}{4}E_0^2L^3=\frac{Q\cdot P}{\omega} 
\end{equation}
being $Q$ the merit factor, $P$ the input power and $V$ the volume of the cavity assumed to be a box of side length $L$. In this case I have to apply the boundary condition
\begin{equation}
    A'(0)=A'(L).
\end{equation}
This yields the modes to be $k_n=2n\pi/L$, being $n$ an integer, and the corresponding frequencies $\omega_n=ck_n$ arising from the Rosen coordinate $v=ct-z$. 

\subsection{Light propagation}

We assume that a beam of light is moving through the box containing one of the modes described above as the cavity is fed through some source of power $P$. There is no electromagnetic interaction between these two electromagnetic fields because light has not self-interaction besides a small effect, dubbed Delbr\"uck scattering, that can be analyzed in quantum electrodynamics and is fourth order. This does not apply here. The propagation of the beam inside the cavity is described by the wave equation
\begin{equation}
     L^2(v)\left(\frac{\partial^2\psi}{\partial x^2}+\frac{\partial^2\psi}{\partial y^2}\right)-4\frac{\partial^2\psi}{\partial u\partial v}=0
\end{equation}
and one sees that the altered geometry by the mode of the cavity can couple it with the laser beam. This equation can be solved by separation of variables setting
\begin{equation}
     \psi(x,y,u,v)={\cal E}(x,y)\phi(u,v)
\end{equation}
being ${\cal E}(x,y)$ an envelope of the beam. This yields the equation for $\phi(u,v)$
\begin{equation}
    -4\frac{\partial^2\phi}{\partial u\partial v}=k^2L^2(v)\phi
\end{equation}
that is
\begin{equation}
    \frac{1}{c^2}\frac{\partial^2\phi}{\partial t^2}-\frac{\partial^2\phi}{\partial z^2}=k^2L^2(ct-z)\phi.
\end{equation}
One can consider $L(ct-z)$ a small quantity and do some perturbation theory yielding
\begin{equation}
    \phi(z,t)\approx\phi_0(z,t)+\frac{ck^2}{2}\int dz'dt'\theta(c(t-t')-(z-z'))L^2(ct'-z')\phi_0(z',t')
\end{equation}
being $\theta(z)$ the Heaviside step function and $\phi_0(z,t)$ the laser beam entering the cavity. Finally, one has
\begin{equation}
\label{eq:sol}
    \psi(x,y,z,t)\approx\psi_0(x,y,z,t)+\frac{ck^2}{2}\int dz'dt'\theta(c(t-t')-(z-z'))L^2(ct'-z')\psi_0(x,y,z',t').
\end{equation}
One sees that there is an additional component to the laser field exiting the cavity that interacts with the mode inside. This can have terms with the frequency shifted and is a purely gravitational effect. In order to see this just note that
\begin{equation}
   L^2(ct-z)=\alpha^2\frac{E_0^2}{2c^2}\left(2+e^{i\omega(t-z/c)}+e^{-i\omega(t-z/c)}\right)
\end{equation}
and, for the laser field,
\begin{equation}
   \psi_0(x,y,z,t)=A(x,y,z)e^{i\omega_Lt}+A^*(x,y,z)e^{-i\omega_Lt}.
\end{equation}
Putting this into eq.(\ref{eq:sol}) one sees that the additional components contribute as
\begin{equation}
   \psi(x,y,z,t)\approx\psi_0(x,y,z,t)+k^2\alpha^2\frac{E_0^2}{4c}\left(A_1(x,y,z)e^{i\omega_Lt}+A_2(x,y,z)e^{i(\omega-\omega_L)t}+A_3(x,y,z)e^{i(\omega+\omega_L)t}+c.c.\right).
\end{equation}
One should observe satellite lines due to the modified geometry of space-time originating from the field inside the cavity. Note also the dependence on $k$ that for a laser can be very large and one gets an overall noticeable effect. This result is important as it shows how, in a small setup and with a proper interferometric device, one could detect very small gravitational effects arising from the effect of an electromagnetic field in a localized region of space-time. It is interesting to note that a dielectric properly inserted into the cavity can enhance this effect significantly changing $\epsilon_0$ into $\epsilon=\epsilon_0\epsilon_r$. For some polymers $\epsilon_r$ can be of the order of $10^5$ \cite{Pohl} or also higher for some ceramic material \cite{Lebey}.

\section{Geometry of resonant cavities and space-time effects \label{secIII}}

We assume now a geometry with a rotational symmetry along the z axis. The geometrical form of the resonant cavity is asymmetrical with respect to two parallel planes being a truncated cone.

\subsection{Modes}

If the resonant cavity has the form of a truncated cone, the modes inside take the form \cite{Minotti:2013mxa}
\begin{eqnarray}
\mathbf{B} &=&-U_0R\left( r\right) S^{\prime }\left( \theta \right) \cos
\left( \omega t\right) \mathbf{e}_{\varphi },  \label{bmode} \\
\mathbf{E}/c &=&U_0\left\{ \frac{R\left( r\right) }{r}n\left( n+1\right)
S\left( \theta \right) \mathbf{e}_{r}\right.  \nonumber \\
&&\left. +\left[ \frac{R\left( r\right) }{r}+R^{\prime }\left( r\right) %
\right] S^{\prime }\left( \theta \right) \mathbf{e}_{\theta }\right\} \sin
\left( \omega t\right)  \label{emode}
\end{eqnarray}%
where $U_0$ is a global constant dependent on the source supplying the cavity and the characteristics of the cavity itself. The functions $R$ and $S$ are defined as 
\begin{eqnarray*}
S\left( \theta \right) &=&P_{n}\left( \cos \theta \right) , \\
R\left( r\right) &=&R_{+}\left( r\right) \cos \alpha +R_{-}\left( r\right)
\sin \alpha , \\
R_{\pm }\left( r\right) &=&\frac{J_{\pm \left( n+1/2\right) }\left(
kr\right) }{\sqrt{kr}},
\end{eqnarray*}%
where $P_{n}$ is the Legendre polynomial of order $n$, $J_{m}$ the Bessel function of the first kind of order $m$, and $\alpha $ and $k$ constants to
be determined along with the order $n$. By boundary conditions, the order $n$ of the Legendre polynomial must satisfy 
\begin{equation}
P_{n}\left( \cos \theta _{0}\right) =0, 
\end{equation}
being $\theta_0$ the semi-angle of the cone, the wavenumber $k$ the condition%
\begin{equation}
\left[ \frac{R_{+}}{r}+R_{+}^{\prime }\right] _{r_{2}}\left[ \frac{R_{-}}{r}%
+R_{-}^{\prime }\right] _{r_{1}}=\left[ \frac{R_{+}}{r}+R_{+}^{\prime }%
\right] _{r_{1}}\left[ \frac{R_{-}}{r}+R_{-}^{\prime }\right] _{r_{2}}, 
\end{equation}
and $\alpha $%
\begin{equation}
\tan \alpha =-\frac{R_{+}\left( r_{2}\right) /r_{2}+R_{+}^{\prime }\left(
r_{2}\right) }{R_{-}\left( r_{2}\right) /r_{2}+R_{-}^{\prime }\left(
r_{2}\right) }. 
\end{equation}
The resonant mode angular frequency is thus determined as $\omega =kc$. From this we can compute the non-zero components of the energy-momentum tensor of the electromagnetic field. We get
\begin{eqnarray}
    F_{10}&=&-F_{01}=U_0 \frac{R\left( r\right) }{r}n\left( n+1\right)S\left( \theta \right) \sin
\left( \omega t\right) \nonumber \\
		F_{20}&=&-F_{02}=\left[ \frac{R\left( r\right) }{r}+R^{\prime }\left( r\right) %
\right] S^{\prime }\left( \theta \right) \sin
\left( \omega t\right) \nonumber \\
    F_{32}&=&-F_{23}=-U_0kR\left( r\right) S^{\prime }\left( \theta \right) \cos
\left( \omega t\right)
\end{eqnarray}
The constant $U_0$ can be obtained using the formula \cite{Minotti:2013mxa}
\begin{equation}
\label{eq:U0}
\frac{\int \left\langle B^{2}\right\rangle dV}{%
\mu _{0}}=\frac{U_0^{2}}{2\mu _{0}}\int \left[ R\left( r\right) S^{\prime
}\left( \theta \right) \right] ^{2}dV=\frac{QP}{\omega },
\end{equation}%
being $Q$ the quality factor of the cavity, $P$ the input power and a time average is applied.

\subsection{Solution of Einstein-Maxwell equations in wave-like form}

It is not difficult to realize that the quantity
\begin{equation}
    \kappa=\frac{8\pi G}{c^4}\approx 2.0765\cdot 10^{-43} N^{-1}
\end{equation}
is small and so we have to eventually apply the linearized theory. 
The equations in the Donder gauge are \cite{Thorne:1980ru}
\begin{equation}
-\frac{1}{2}\Box\overline{h}_{\mu \nu }=%
\kappa (T_{\mu \nu }+\tau_{\mu\nu}) \label{Gik0}
\end{equation}
being $\tau_{\mu\nu}$ the gravity stress-energy tensor and
\begin{equation}
  \overline{h}_{\mu \nu }\equiv h_{\mu \nu }-\frac{1}{2}h\eta _{\mu \nu }.
\end{equation}
We have set at the start
\begin{equation}
    g_{\mu\nu}=\eta_{\mu\nu}+h_{\mu\nu}
\end{equation}
being $\eta_{\mu\nu}$ the flat metric and $h_{\mu\nu}$ the gravity field. This is not a tensor but it is not a concern here. We work out the analysis as given in \cite{Wein}. We get the general solution
\begin{equation}
   \overline{h}_{\mu \nu }({\bm x},t)=-2\kappa\int_Vd^3x'\frac{(T_{\mu \nu }
	+\tau_{\mu\nu})\left({\bm x}',t-\frac{{\bm x}-{\bm x}'}{c}\right)}{|{\bm x}-{\bm x}'|}
\end{equation}
being $\tau_{\mu\nu}$ the Landau-Lifshitz pseudotensor of the gravity field. We introduce the constant
\begin{equation}
    l_0^{-2} = 2\kappa\frac{U_0^2}{\mu_0}\approx 3.3\cdot 10^{-37} U_0^2
\end{equation}
with $U_0^2$ given in $T^2$ and being the definition of a length. This means that
\begin{equation}
   \overline{h}_{\mu \nu }({\bm x},t)=-l_0^{-2}\int_V d^3x'\frac{(\overline T_{\mu \nu }
	+\mu_0U_0^{-2}\tau_{\mu\nu})\left({\bm x}',t-\frac{{\bm x}-{\bm x}'}{c}\right)}{|{\bm x}-{\bm x}'|}
\end{equation}
being $\overline T_{\mu \nu }$ the dimensionless energy-momentum tensor of the electromagnetic field inside the cavity. We can remove $l_0$ by changing the length scale in the integral and obtain
\begin{equation}
   \overline{h}_{\mu \nu }(\overline{\bm x},\overline t)=-\int_V d^3\overline x'\frac{(\overline T_{\mu \nu }
	+2\kappa\overline\tau_{\mu\nu})\left(\overline {\bm x}',\overline t-\frac{\overline{\bm x}-\overline{\bm x}'}{c}\right)}{|\overline{\bm x}-\overline{\bm x}'|}
\end{equation}
having set $\overline{\bm x}={\bm x}/l_0$ and $\overline t = t/(l_0/c)$. $\overline\tau_{\mu\nu}$ is the normalized gravity pseudotensor. This has a prefactor $(2\kappa)^{-1}$. $l_0$ is really large unless we are in the field of a magnetar. Then, the integral is easy to evaluate to give
\begin{equation}
\label{eq:hmunu}
   \overline{h}_{\mu \nu }(\overline{\bm x},\overline t)=-
	L(\overline{\bm x})\left(\overline T_{\mu \nu }
	+2\kappa\overline\tau_{\mu\nu}\right)\left(\overline {\bm x},\overline t\right)
\end{equation}
being
\begin{equation}
    L(\overline{\bm x})=\int_V d^3\overline x'\frac{1}{|\overline{\bm x}-\overline{\bm x}'|}
\end{equation}
a geometrical factor obtained by integrating on the volume of the frustum. Eq.(\ref{eq:hmunu}) would be a differential equation for $\overline{h}_{\mu \nu }$ but, in a first approximation, we can assume that the derivatives of it are negligible and we are left with the result
\begin{equation}
\label{eq:hmunu_f}
   \overline{h}_{\mu \nu }(\overline{\bm x},\overline t)=-
	L(\overline{\bm x})\overline T_{\mu \nu }\left(\overline{\bm x},\overline t\right).
\end{equation}
This is our key result and can be stated in the same way as inductance enters into electromagnetic field.

\subsection{Gravitational susceptibility}

The susceptibility of the frustum can be evaluated by computing the integral, in cylindrical coordinates,
\begin{equation}
   L(r,z,\theta)=\int_0^hdz'\int_0^{2\pi}d\theta'\int_0^{\frac{r_2-r_1}{h}z'+r_2}r'dr'\frac{1}{\sqrt{r^2+r'^2+(z-z')^2-2rr'\cos(\theta-\theta')}}
\end{equation}
that is rather involved. A way out is to note that
\begin{equation}
   \Delta_2\frac{1}{|{\bm x}-{\bm x}'|}=\delta^3({\bm x}-{\bm x}')
\end{equation}
and so
\begin{equation}
   \Delta_2L(\overline{\bm x})=1.
\end{equation}
The solution of this equation is
\begin{equation}
    L(\overline{\bm x})=L_o(\overline{\bm x})+a+b\ln(\overline r)+\frac{\overline r^2}{4}
\end{equation}
being $L_o(\overline{\bm x})$ a solution of the equation $\Delta_2L_o(\overline{\bm x})=0$, we assume it to be zero, and we have to set the condition for the frustum
\begin{equation}
\label{eq:rdiz}
    \overline r(\overline z)=\frac{r_2-r_1}{h}\overline z+\frac{r_1}{l_0}.
\end{equation}
This yields
\begin{eqnarray}
\label{eq:a&b}
a&=&\ln^{-1}\frac{r_2}{r_1}\left(\frac{1}{4}\frac{r_2^2}{l_0^2}\ln\frac{r_1}{l_0}
-\frac{1}{4}\frac{r_1^2}{l_0^2}\ln\frac{r_2}{l_0}\right) \nonumber \\
b&=&\ln^{-1}\frac{r_2}{r_1}\left(\frac{1}{4}\frac{r_1^2}{l_0^2}
-\frac{1}{4}\frac{r_2^2}{l_0^2}\right).
\end{eqnarray}
These equations appear rather interesting as, by a proper choice of parameters, one can make a gravitational effect more or less relevant in the physics of the problem. 
The case of interest is that implying the minor radius $r_1$ going to 0 and the major radius $r_2$ increasing to infinity approaching a cone. In this case the geometry can help to alleviate the smallness of the ratio $G/c^4$ turning this into an observable effect. 

\subsection{Gravitational reaction}
 
In order to evaluate the contribution of the gravitational field to momentum one has to evaluate the approximate Landau-Lifshitz tensor \cite{LL2}
\begin{equation}
    \tau^{\alpha\beta}_{LL}=\frac{c^4}{16\pi G}\langle h_\mu^{\nu,\alpha}h_\nu^{\mu,\beta}\rangle
\end{equation}
where the average $\langle\cdots\rangle$ is intended on time and on the $\cos\theta$. The comma means ordinary derivative $\partial_\alpha$. In order to evaluate this correction we observe that, for the tensor $F^{\mu\nu}$ is
\begin{eqnarray}
   &&F^{01}=-F^{10}=E_r, \qquad F^{02}=-F^{20}=E_\theta, \nonumber \\
	 &&F^{23}=-F^{32}=B_\phi
\end{eqnarray}
and all other components are zero. So, in vacuum,
\begin{eqnarray}
   &&T^{00}=\frac{1}{2}\left(\epsilon_0E_r^2+\epsilon_0E_\theta^2+\frac{B_\phi^2}{\mu_0}\right), \qquad T^{01}=T^{10}=\frac{E_\theta B_\phi}{\mu_0}, 
	\qquad T^{02}=T^{20}=-\frac{E_rB_\phi}{\mu_0}, \nonumber \\
	 &&T^{11}=\frac{1}{2}\epsilon_0E_r^2-\frac{1}{2}\left(\epsilon_0E_\theta^2+\frac{B_\phi^2}{\mu_0}\right),
	\qquad T^{22}=\frac{1}{2}\epsilon_0E_\theta^2-\frac{1}{2}\left(\epsilon_0E_r^2+\frac{B_\phi^2}{\mu_0}\right), \nonumber \\
	 &&T^{33}=\frac{1}{2}\frac{B_\phi^2}{\mu_0}-\frac{1}{2}\left(\epsilon_0E_r^2+\epsilon_0E_\theta^2\right), \qquad T^{12}=T^{21}=\epsilon_0E_rE_\theta.
\end{eqnarray}
In a first approximation it is
\begin{equation}
    h_\mu^{\nu,\alpha}\approx L^{,\alpha}\eta^{\nu\delta}T_{\mu\delta}
\end{equation}
then
\begin{equation}
    \tau^{\alpha\beta}_{LL}\approx\frac{c^4}{16\pi G}\langle L^{,\alpha}L^{,\beta}\eta^{\nu\delta}\eta^{\gamma\mu}T_{\mu\delta}T_{\gamma\nu}\rangle.
\end{equation}
Now, we observe that $r$ in $L$ depends on $z$ through the equation of the truncated cone (\ref{eq:rdiz}) and so, the only term that survives of this is
\begin{equation}
    \tau^{zz}_{LL}\approx\frac{c^4}{16\pi G}\langle L^{,z}(r(z))L^{,z}(r(z))\eta^{\nu\delta}\eta^{\gamma\mu}T_{\mu\delta}T_{\gamma\nu}\rangle.
\end{equation}
It is important to observe that now we have the square of the energy-momentum tensor and so the average in time is different from zero granting the presence of a force along $z$ axis. All the components of the electromagnetic field inside the frustum concur to the force along $z$. We assume the energy-momentum tensor normalized in unit of $U_0$. Then, the force has an overall factor of $U_0^4$ against the really tiny contribution of the $G/c^4$ factor of the gravitational field. In this way we can evaluate the force on the walls along z axis as
\begin{eqnarray}
     F_{z1}&=&\pi r_1^2\frac{1}{2\cos\theta_0}\int_{-\cos\theta_0}^{\cos\theta_0}d\cos\theta\tau_{LL}^{zz}=
		\frac{1}{2\cos\theta_0}\int_{-\cos\theta_0}^{\cos\theta_0}d\cos\theta\frac{c^4}{16\pi G}
		\langle h_\mu^{\nu,z}h_\nu^{\mu,z}\rangle= \nonumber \\
		&&\frac{c^4}{16\pi G}\frac{1}{2\cos\theta_0}\int_{-\cos\theta_0}^{\cos\theta_0}d\cos\theta\left.
		\left[L'(r)\frac{dr}{dz}\right]^2\right|_{r=r_1}\langle {\overline T}_{\mu}^{\nu}{\overline T}_{\nu}^{\mu}\rangle
\end{eqnarray}
and
\begin{equation}
     F_{z2}=\pi r_2^2\frac{c^4}{16\pi G}\frac{1}{2}\int_{-\cos\theta_0}^{\cos\theta_0}d\cos\theta\left.
		\left[L'(r)\frac{dr}{dz}\right]^2\right|_{r=r_2}\langle {\overline T}_{\mu}^{\nu}{\overline T}_{\nu}^{\mu}\rangle
\end{equation}
being $h$ the height of the frustum. The resulting thrust is given by $T=F_{z_2}-F_{z_1}$. We note that $dS_k=\pi r_k^2d\cos\theta$ with $k=1,2$ for the two opposite faces of the frustum. Here $\theta$ runs from $-\theta_0$ to $\theta_0$ assuming $2\theta_0$ the full opening angle of the frustum. One has,
\begin{eqnarray}
     F_{z_k}&=&
		\pi r_k^2\frac{c^4}{16\pi G}\left[\left(\frac{b}{r_k}+\frac{r_k}{2l_0^2}\right)\frac{r_2-r_1}{h}\right]^2\frac{1}{2\cos\theta_0}
		\int_{-\cos\theta_0}^{\cos\theta_0}d\cos\theta\times \nonumber \\
		&&\frac{1}{T}\int_0^Tdt {\overline T}_{\mu}^{\nu}\left(\sqrt{r_k^2+h^2\delta_{k,2}},\theta,t\right){\overline T}_{\nu}^{\mu}\left(\sqrt{r_k^2+h^2\delta_{k,2}},\theta,t\right).
\end{eqnarray}
In our case it is
\begin{equation}
\label{eq:T^2}
    {\overline T}_{\mu}^{\nu}{\overline T}_{\nu}^{\mu}=
		({\overline T}_0^0)^2+({\overline T}_1^1)^2+({\overline T}_2^2)^2+({\overline T}_3^3)^2+2({\overline T}_0^1)^2
		+2({\overline T}_0^2)^2+2({\overline T}_1^2)^2
\end{equation}
and the only integrals to be evaluated in time are
\begin{eqnarray}
    &&\frac{1}{T}\int_0^Tdt\sin^2\omega t\cos^2\omega t=\frac{1}{8} \nonumber \\
		&&\frac{1}{T}\int_0^Tdt\sin^4\omega t=\frac{1}{T}\int_0^Tdt\cos^4\omega t=\frac{3}{8}
\end{eqnarray}
that are not zero. There is a net thrust. Now, let us evaluate the order of magnitude of it. In order to do this we assume that each single term gives a similar contribution to to sum and we content ourselves with the integral
\begin{equation}
    {\cal T}_k=\frac{1}{2\cos\theta_0}\int_{-\cos\theta_0}^{\cos\theta_0}d\cos\theta
		\langle({\overline T}_0^0)^2\left(\sqrt{r_k^2+h^2\delta_{k,2}},\theta,t\right)\rangle
\end{equation}
that yields
\begin{equation}
    {\cal T}_k=\frac{1}{2\cos\theta_0}\int_{-\cos\theta_0}^{\cos\theta_0}d\cos\theta\frac{1}{4}\left(\frac{3}{8}{\overline E}_r^4+
		\frac{3}{8}{\overline E}_\theta^4+\frac{3}{8}{\overline B}_\phi^4+\frac{3}{4}{\overline E}_r^2{\overline E}_\theta^2
		+\frac{1}{4}{\overline E}_r^2{\overline B}_\phi^2
		+\frac{1}{4}{\overline E}_\theta^2{\overline B}_\phi^2\right).
\end{equation}
Overall we get a numerical factor depending on $\theta_0$, $r_1$ or $r_2$ and $h$. So, finally
\begin{equation}
\label{eq:Fz}
    F_{z_k}=\frac{c^4}{16G}\left[\left(\frac{b}{r_k}+
		\frac{r_k}{2l_0^2}\right)\frac{r_2-r_1}{h}\right]^2{\cal G}_k(r_k,h\delta_{k,2},\theta_0)
\end{equation}
being ${\cal G}_k(r_k,h\delta_{k,2},\theta_0)$ is a purely geometrical factor arising from the distribution of the electromagnetic field inside the cavity.

\subsection{Estimation of the force}

The numbers involved in this kind of computation, as usual for gravity at this level, are really small. So, we need to understand what is the better geometrical form to choose to get an effect at least amenable to observation in current laboratory experiments. We see that in eq.(\ref{eq:Fz}) enters a multiplicative constant and two factors crucially depending on the geometry of the resonant cavity. Let us evaluate each one of these. We define
\begin{equation}
    L_1(r)=\frac{c^4}{16G}\left[\left(\frac{b}{r}+\frac{r}{2l_0^2}\right)\frac{r_2-r_1}{h}\right]^2.
\end{equation}
It is easy to see that $L_1(r)\rightarrow 0$ as $r_1\rightarrow r_2$ and so, we have to exclude a cylindrical geometry. On the other side, when $r_2\gg r_1$ we get
\begin{equation}
    L_1(r)\approx\frac{\pi^2G}{c^4}\frac{r_2^6}{r^2h^2\ln^2\left(\frac{r_2}{r_1}\right)}\frac{U_0^4}{\mu_0^2}
\end{equation}
and this geometrical factor can compensate largely the ratio $G/c^4$ producing an overall macroscopic effect. Also note that, from eq.(\ref{eq:U0}),
\begin{equation}
    \frac{U_0^{2}}{2\mu _{0}}C(k\cdot r_1,k\cdot r_2,\theta_0)=\frac{QPk^2}{c}
\end{equation}
and then $U_0^4/\mu_0^2\propto Q^2P^2k^4/c^2$ so, higher are $Q$ of cavity and the input power and higher is the force observed. This can also be rewritten as
\begin{equation}
    c^2\epsilon_0\frac{U_0^{2}}{2}C(k\cdot r_1,k\cdot r_2,\theta_0)=\frac{QPk^2}{c}
\end{equation}
where use has been made of the formula $c^2=1/\epsilon_0\mu_0$. The dependence on materials enters here changing $\epsilon_0\rightarrow\epsilon_r\epsilon_0$ and $\mu_0\rightarrow\mu_r\mu_0$.

Turning the attention to the contribution ${\cal G}_k$, we note that there is a numerical factor due to the average of the fourth power of the given Legendre polynomial or its derivative. This is just a numerical factor, to be evaluated case by case, that depends on $\theta_0$. Then terms that contribute are
\begin{eqnarray}
     {\overline E}^4_r(r_n,\theta_0)&=&\frac{R^4\left( k\cdot r_n\right) }{k^4\cdot r_n^4}n^4\left(n+1\right)^4\langle \left[S\left( \theta \right)\right]^4\rangle_{\theta_0} \nonumber \\
		 {\overline E}^4_\theta(r_n,\theta_0)&=&\left[\frac{R\left( k\cdot r_n\right) }{k\cdot r_n}+R^{\prime }\left( k\cdot r_n\right) \right]^4
		 \langle \left[S^{\prime }\left( \theta \right)\right]^4\rangle_{\theta_0} \nonumber \\
		 {\overline B}^4_\phi(r_n,\theta_0)&=&R^4\left( k\cdot r_n\right) \langle\left[S^{\prime }\left( \theta \right)\right]^4\rangle_{\theta_0} \nonumber \\
		{\overline E}^2_r(r_n,\theta_0){\overline B}^2_\phi(r_n,\theta_0) &=&
		\frac{R^2\left( k\cdot r_n\right) }{k^2\cdot r_n^2}n^2\left(n+1\right)^2
		R^2\left( k\cdot r_n\right) \langle\left[S\left( \theta \right)\right]^2\left[S^{\prime }\left( \theta \right)\right]^2\rangle_{\theta_0} \nonumber \\
		{\overline E}^2_\theta(r_n,\theta_0){\overline B}^2_\phi(r_n,\theta_0) &=&
		\left[\frac{R\left( k\cdot r_n\right) }{k\cdot r_n}+R^{\prime }\left( k\cdot r_n\right) \right]^2
		R^2\left( k\cdot r_n\right) \langle\left[S^{\prime }\left( \theta \right)\right]^4\rangle_{\theta_0}
\end{eqnarray}
with $n=1,2$ and we gave explicitly the dependence on $k$. We note that, for $k\cdot r_2\gg 1$, the only relevant component is the fourth power of the magnetic field. So, one has the final formula
\begin{equation}
    {\cal G}_k(r_k,h\delta_{k,2},\theta_0)\approx
		R^4\left( k\cdot r_n\right) \langle\left[S^{\prime }\left( \theta \right)\right]^4\rangle_{\theta_0}.
\end{equation}
Finally, we can get the direction of the thrust on the cavity. We get
\begin{equation}
     T=F_{z_2}-F_{z_1}=\frac{\pi^2G}{c^4}\frac{r_2^6}{h^2\ln^2\left(\frac{r_2}{r_1}\right)}\frac{U_0^4}{\mu_0^2}{\cal G}_k(\sqrt{r_2^2+h^2},h,\theta_0)-
		\frac{\pi^2G}{c^4}\frac{r_2^6}{h^2\ln^2\left(\frac{r_2}{r_1}\right)}\frac{U_0^4}{\mu_0^2}{\cal G}_k(r_1,0,\theta_0),
\end{equation}
provided $r_2\gg r_1$ and $r_1$ small enough. This can also be written, using the formula $c^2=1/\epsilon_0\mu_0$, as
\begin{equation}
     T=F_{z_2}-F_{z_1}=\pi^2G\epsilon_0^2U_0^4\frac{r_2^6}{h^2\ln^2\left(\frac{r_2}{r_1}\right)}\left[{\cal G}_k(\sqrt{r_2^2+h^2},h,\theta_0)-
		{\cal G}_k(r_1,0,\theta_0)\right].
\end{equation}
This is our final result that shows how the interplay between electromagnetic field, space-time geometry and geometry of a resonant cavity can yield a finite thrust to the cavity itself without violating any law of physics. When a dielectric is inserted into the cavity, the above formula should change with $\epsilon_0\rightarrow\epsilon$ and $U_0$ should change accordingly for $\mu$. One has
\begin{equation}
    \frac{U_0^{2}}{2}C(k\cdot r_1,k\cdot r_2,\theta_0)=\mu _{0}\mu_r\sqrt{\epsilon_r\mu_r}\frac{QPk^2}{c}
\end{equation}

This effect appears to be really small. If we assume a cavity with $r_1=0.025\ m$, $r_2=0.1\ m$ and $h=0.1\ m$, for a mode having $N=42.272946$, $k=701.897366\ m^{-1}$, $\alpha=0$ and $\nu=210.423537\ GHz$ the given thrust is $T=5.945503835\cdot 10^{-22}\ N$, if we assume a dielectric with $\mu_r\approx 1$ and $\epsilon_r\approx 10^5$ (e.g. a conjugated polymer \cite{Pohl,Lebey}) is inserted into the cavity. For this effect the scale is set by the constant
\begin{equation}
   T_0=\pi^2G\epsilon^2U_0^4r_2^4\approx 5\cdot 10^{-32}\epsilon_r^2U_0^4r_2^4\ N.
\end{equation}
Such an effect could be evidenced with interferometric techniques by shooting a laser beam inside the cavity in a Mach-Zender interferometer at small input powers.

\section{Brans-Dicke-Maxwell theory}

The effect due to Einstein theory is really miniscule due to the ratio $G/c^4$. Things could be somewhat different if the theory that describes gravity is a Brans-Dicke. In this case the gravitational constant $G$ is changed into a $1/\phi$ being $\phi$ a scalar-field. The coupling with the electromagnetic field gives a set of equations that can be solved in some cases \cite{Ciftci:2015cua,Baykal:2009vs,Baykal:2009zz,Cai:1996pj}. The special case of four dimensions implies that, being the energy-momentum tensor of the electromagnetic field traceless, the scalar field does not couple directly with the electromagnetic field and we have the equation to solve
\begin{equation}
    (-g)^{-1/2}\partial_\mu(-g)^{1/2}g^{\mu\nu}\partial^\nu\Phi = 0.
\end{equation}
In our computation we derived
\begin{equation}
    \overline{h}_{\mu \nu }(\overline{\bm x},\overline t)=-
	L(\overline{\bm x})\overline T_{\mu \nu }\left(\overline{\bm x},\overline t\right).
\end{equation}
It is important to note that also $\overline{h}_{\mu \nu }$ is traceless and then, $\overline{h}_{\mu \nu }=h_{\mu\nu}$. Finally,
\begin{equation}
    g_{\mu\nu}=\eta_{\mu\nu}-L(\overline{\bm x})\overline T_{\mu \nu }\left(\overline{\bm x},\overline t\right).
\end{equation}
Now,
\begin{equation}
    \operatorname{det}(-g)=\operatorname{det}(-\eta-h)=1+\frac{1}{2}h^\mu_\nu h^\nu_\mu+O(h^3)
\end{equation}
as $\operatorname{Tr}(h)=0$ and we used the formula $\operatorname{det}(I+\epsilon A)=\sum_{n=0}^\infty (\epsilon^n\operatorname{Tr}(A^n))/n!$ considering $h$ a small correction. Then, stopping at second order, we get the equation for the scalar field
\begin{equation}
    -\partial^2\Phi+\partial_\mu (h^{\mu\nu}\partial_\nu\Phi)-\frac{1}{2}\partial_\mu (h^\rho_\lambda h^\lambda_\rho\partial^\mu\Phi) = 0.
\end{equation}
This equation takes the form
\begin{equation}
    -\partial^2\Phi-\partial_\mu (L(\overline{\bm x})\overline T^{\mu \nu }\partial_\nu\Phi)
		-\frac{1}{2}\partial_\mu (L^2(\overline{\bm x})\overline T^\rho_\lambda \overline T^\lambda_\rho\partial^\mu\Phi) = 0.
\end{equation}
In order to get an estimation of the gravitational constant in the cavity, we take an average value for $L(r)$ and call it $L_0$. Then, we consider a mode $k_\mu$ and use the conservation of energy-momentum tensor $k_\mu T^{\mu\nu}=0$ and introduce a new variable $\xi=k\cdot x$. This yields
\begin{equation}
   \frac{d^2\Phi}{d\xi^2}+\frac{1}{2}\frac{d}{d\xi} (L_0^2\overline T^\rho_\lambda \overline T^\lambda_\rho\frac{d\Phi}{d\xi}) = 0.
\end{equation}
Such an equation can be cast into an integral form noting that $\overline T^\rho_\lambda\rightarrow{\tilde T}^\rho_\lambda/\Phi$. The, the equation can be cast into an integral form as
\begin{equation}
    \Phi(\xi)=\frac{1}{G(kr_2-kr_1)}\int_{kr_1}^\xi\frac{d\xi'}{1+\frac{1}{2}L_0^2\Phi^2(\xi')\tilde T^\rho_\lambda(\xi') \tilde T^\lambda_\rho(\xi')}
\end{equation}
where we have assumed that outside the cavity the Newton constant must be recovered and we have integrated on the volume of the cavity.
This is a consistency equation that can be solved by iterating on $\Phi$ and assuming, at the leading order, e.g. $\Phi=1/G$. Then,
\begin{equation}
    \Phi(\xi)=\frac{1}{G(kr_2-kr_1)}\int_{kr_1}^\xi\frac{d\xi'}{1+\frac{1}{2}L_0^2\overline T^\rho_\lambda(\xi') \overline T^\lambda_\rho(\xi')}
\end{equation}
We conclude that, if a Brans-Dicke model holds, the gravitational constant inside the cavity is modified by the electromagnetic field in it.

\section{Conclusions \label{secIV}}

I have shown how a plane wave could produce a gravitational effect inside a cavity that could be observed using a propagating laser beam inside it. The effect could be unveiled using an interferometer or observing the components of the laser field outside the cavity. Components with a shifted frequency, due to the modes inside the cavity, should be seen. This could explain some recent results with interferometric setup obtained at NASA with a resonator having the form of a box. A local warp of the geometry due to the electromagnetic field pumped inside the cavity could be a satisfactory explanation. From a physical standpoint this could be a really breakthrough paving the way to table-top experiments in general relativity and marking the starting point of space-time engineering.

Then, I considered a frustum in the form of a truncated cone. I have shown that general relativity introduce a large scale that makes all the effects really miniscule. 
For the frustum I have shown that the gravitational effects can be described by a susceptibility multiplying the energy-momentum tensor of the electromagnetic field inside the cavity. Due to this particular geometry, it can be shown that the susceptibility can be made significant by a proper choice of the geometrical parameters of the cavity yielding thrust without violating any law of physics. 
This effect could amenable to observation with a proper interferometric setup.

Finally, if the model of gravity that holds is described by a Brans-Dicke theory, inside the cavity the electromagnetic field induces a modification of the gravitational constant. 

\begin{acknowledgments}
I would like to thank Jose Rodal for a significant exchange of points of view on these results that made possible a deeper understanding of them in aerospace applications.
\end{acknowledgments}

\end{document}